\documentclass[aps,prd,twocolumn,eqsecnum,showpacs,floatfix]{revtex4}
\usepackage[dvips]{color,graphicx}
\usepackage{subfigure}
\usepackage{amsfonts}
\usepackage{amssymb}

\newcommand{\dalm}{\kern1pt\vbox{\hrule height 0.9pt\hbox{\vrule width
0.9pt\hskip 2.5pt\vbox{\vskip 5.5pt}\hskip 3pt\vrule width 0.3pt}\hrule height
0.3pt}\kern1pt}

\begin{document}


\title{
Self-similar cosmological solutions with dark energy. II: \\
black holes, naked singularities and wormholes
}

\author{$^{1,2,3,4}$Hideki Maeda\footnote{Electronic
address:hideki@cecs.cl},
$^{4}$Tomohiro Harada\footnote{Electronic
address:harada@rikkyo.ac.jp}, and $^{5,6}$B.~J.~Carr\footnote{Electronic address:B.J.Carr@qmul.ac.uk}}
\affiliation{
$^{1}$Centro de Estudios Cient\'{\i}ficos (CECS), Arturo Prat 514, Valdivia, Chile\\
$^{2}$Department of Physics, International Christian University, 3-10-2 Osawa, Mitaka-shi, Tokyo 181-8585, Japan\\
$^{3}$Graduate School of Science and Engineering, Waseda University, Tokyo 169-8555, Japan\\
$^{4}$Department of Physics, Rikkyo University, Tokyo 171-8501, Japan\\
$^{5}$Astronomy Unit, Queen Mary, University of London, Mile End Road, London E1 4NS, UK\\
$^{6}$Research Center for the Early Universe, Graduate School of Science, University of Tokyo, Tokyo 113-0033, Japan}
\date{\today}

\begin{abstract}                
We use a combination of numerical and analytical methods, exploiting the equations derived in a preceding paper, to classify all spherically symmetric self-similar solutions which are asymptotically Friedmann at large distances and contain a perfect fluid with equation of state $p=(\gamma -1)\mu$ with $0<\gamma<2/3$. 
The expansion of the Friedmann universe is accelerated in this case.
We find a one-parameter family of self-similar solutions representing a
 black hole embedded in a Friedmann background.
This suggests that, in contrast to the 
positive pressure case,
black holes in a universe with dark energy can grow as fast as the Hubble horizon if they are not too large. There are also self-similar solutions which contain a central naked singularity with negative mass and solutions which represent a Friedmann universe connected to either another Friedmann universe or some other cosmological model.
The latter are interpreted as self-similar cosmological white hole or wormhole solutions.
The throats of these wormholes are defined as two-dimensional spheres with minimal area on a spacelike hypersurface and they are all non-traversable because of the absence of a past null infinity.
\end{abstract}


\maketitle

\section{Introduction} 
The study of black holes in stationary and asymptotically flat spacetimes 
has led to many remarkable insights, such as 
the ``no hair'' theorem and the discovery of the thermodynamic properties of black holes.
In contrast, black holes in a Friedmann background have not been fully 
investigated and many important open questions remain.
When the size of the black hole is much smaller than the
cosmological horizon, asymptotically flat black hole solutions will
presumably provide a good approximation.
However, for black holes comparable to the horizon size --
such as primordial black holes (PBHs) formed in the early universe~\cite{hawking1971} -- 
one would need to take into account the effects of 
the cosmological expansion.

The main issue addressed in this paper is the growth of such ``cosmological'' black holes.
Zel'dovich and Novikov~\cite{zn1967} first discussed the possibility of self-similar growth for a black hole in a Friedmann universe in 1967. This suggested that the size of a PBH  might always be the same fraction of the size of the particle horizon until the end of the radiation-dominated era, when they would have a mass of order $10^{15}M_{\odot}$. Since there is no evidence that such huge black holes exist, this seemed to imply that PBHs never formed at all. 
More recently, the possibility of self-similar PBH growth in the quintessence scenario has led to the suggestion that PBHs could provide 
seeds for the supermassive black holes thought to reside in galactic nuclei~\cite{bm2002}.
However, both these arguments are based on a simple Newtonian analysis.

The first general relativistic analyses of self-similar cosmological black holes was given by Carr and Hawking~\cite{ch1974} for a radiation fluid in 1974.
They showed that there is no self-similar solution which contains a black hole attached to an exact Friedmann background via a sonic point (i.e. in which the black hole forms by purely causal processes). This result was extended to the perfect fluid case with $p=(\gamma-1)\mu$ and $1 \le \gamma < 2$ by Carr~\cite{carr1976} and Bicknell and Henriksen~\cite{bh1978a}.
For a stiff fluid ($\gamma=2$), it was originally claimed by Lin et al.~\cite{lcf1976} that self-similar cosmological black holes {\it could} exist. However, Bicknell and Henriksen~\cite{bh1978b} subsequently showed that this solution does not represent a black hole after all, since Lin et al. had misidentified the event horizon.
Although Bicknell and Henriksen did
construct some numerical self-similar
solutions, in these the stiff fluid turns into null dust at a timelike hypersurface, which seems physically implausible. Indeed, we have recently proved rigorously the non-existence of self-similar cosmological black holes for both a stiff fluid and a scalar field~\cite{hmc2006}.

In the present paper, we focus on fluids with $0<\gamma<2/3$. 
Although this equation of state violates the strong energy conditions, it may still be very relevant for cosmology -- both in the early universe (when inflation occurred~\cite{linde}) and at the current epoch (when the acceleration may be driven by some form of dark energy~\cite{supernova}). 
Such matter exhibits ``anti-gravity'' -- in the sense that the active gravitational mass in the Raychaudhuri equation is negative -- but the dominant, null and weak energy conditions still hold. Recently, the accretion of dark energy or a phantom field onto a black hole has also been studied~\cite{bde2004,bde2005}, but the cosmological expansion is again neglected in these analyses.

The $0<\gamma<2/3$ case is particularly important in the cosmological
context because there then exists a one-parameter family of self-similar solutions which are
exactly asymptotic to the flat Friedmann model at large distances~\cite{nusser2006}.
In the positive pressure case, the solutions cannot be ``properly'' asymptotic Friedmann because they exhibit a solid angle deficit at infinity, so it would be more accurate to describe them by ``quasi-Friedmann''~\cite{mkm2002}.
In the present paper, we obtain a one-parameter family of 
physically reasonable self-similar cosmological black hole solutions numerically.
This strongly suggests that a black 
hole in a universe filled with dark 
energy {\it can} grow in a self-similar manner.
In ref.~\cite{nusser2006}, this system was also numerically 
investigated but no black hole solution was reported.
This is because one needs to consider the analytic extension of the solutions 
to find the black holes and this was not done in ref.~\cite{nusser2006}.

We also find that there exists a class of self-similar cosmological wormhole solutions.
A wormhole is an object connecting two (or more) infinities.
Einstein's equations certainly permit such solutions. 
For example, the well-known {\it static} wormhole studied by Morris and Thorne~\cite{mt1988} 
connects two asymptotically flat spacetimes.
Although the concept of a wormhole is originally topological and global, it is possible to define it locally by a two-dimensional surface of minimal area on a non-timelike hypersurface. 
However, in any local definition, we inevitably face the problem of time-slicing, because the concept of ``minimal area'' is slice-dependent.

Hochberg and Visser~\cite{hv1997} and Hayward~\cite{hayward1999} 
define a wormhole throat 
on a null hypersurface, in which case,
a light ray can travel from one infinity to another infinity. Their definition excludes the maximally extended Schwarzschild spacetime from the family of wormhole spacetimes. It also implies that in the non-static situation 
the null energy condition ($\mu \ge 0$ for the present matter model) must be violated at the wormhole throat, so the existence of wormholes might seem implausible~\cite{visser,hv1997,hayward1999}.
However, their definition of a wormhole is physically reasonable
only in the presence of a past null infinity, and in the cosmological
situation this often does not apply because there is a big-bang
singularity. Thus, wormholes in their sense may be too restrictive
in the cosmological context.

This motivates us to define a wormhole throat on a {\it spacelike} hypersurface.
Here we adopt the conventional definition of traversability in which an observer can go from one infinity to another distinct infinity.
Our numerical solutions are not wormholes in the sense of Hochberg and Visser~\cite{hv1997} or Hayward~\cite{hayward1999} but they are wormholes in the sense that they connect two infinities. 
Also, although dark energy violates the strong energy condition, the null energy condition is still satisfied in our wormhole solutions.
Numerical solutions containing a super-horizon-size black hole in a Friedmann universe filled with a massless scalar field provide another interesting example of this sort of wormhole~\cite{hc2005c}. 
These involve a wormhole structure with two distinct null infinities, where all possible energy conditions are satisfied.

We find that there are three types of cosmological self-similar 
wormhole solutions. The first connects two
Friedmann universes, the second a Friedmann and quasi-Friedmann universe, the third a Friedmann universe and a quasi-static spacelike infinity. 
In fact, solutions of the third type are naturally interpreted as a Friedmann universe emergent from a white hole.
In order to obtain these results, we utilize the formulation and
asymptotic analyses presented in the accompanying 
paper~\cite{hmc1} (henceforth Paper I).
The combination of these two papers completes 
the classification of asymptotically Friedmann spherically symmetric 
self-similar solutions with dark energy.

The plan of this paper is as follows.
In Section~\ref{sec:numerical_results}, we present the basic variables 
and describe the numerical results.
(Paper I provides the detailed formulation and field equations.)
In Section~\ref{sec:physical_interpretation}, we classify the
self-similar solutions in terms of their physical properties.
Section~\ref{sec:summary_discussion} provides a summary and discussion.

\section{Numerical Results}
\label{sec:numerical_results}

We consider a spherically symmetric self-similar spacetime with a perfect fluid and adopt comoving coordinates.
The line element can be written as
\begin{equation}
ds^2 = -e^{2\Phi(z)}dt^2+e^{2\Psi(z)}dr^2+ r^2 S^2(z) d\Omega^2, \label{metric}
\end{equation}
where $z \equiv r/t$ is the self-similar variable, $R \equiv rS$ is the areal radius, and we adopt units with $c=1$.
The Einstein equations imply that the pressure $p$, energy density $\mu$ and the Misner-Sharp mass $m$ must have the form 
\begin{eqnarray}
8\pi G\mu&=&\frac{W(z)}{r^{2}}, \label{defW}\\
8\pi Gp&=&\frac{P(z)}{r^{2}},\label{defP} \\
2Gm&=&rM(z).\label{defM}
\end{eqnarray}
We assume an equation of state $p=(\gamma-1)\mu$ with $0<\gamma<2/3$.
A crucial role is played by the velocity function $V \equiv |z|e^{\Psi-\Phi}$, which is the velocity of the fluid relative to the similarity surface of constant $z$.
A similarity surface 
is spacelike for $(1-V^2)e^{2\Phi}<0$, timelike for $(1-V^2)e^{2\Phi}>0$, and null for $(1-V^{2})e^{2\Phi}=0$.
In the last case, the surface is called a ``similarity horizon''.  
We also use trapping horizon notation for describing the 
physical properties of solutions.
(See refs.~\cite{hayward1994,hayward1996} for the definition of a trapping horizon.)

The basic field equations are derived from the Einstein equations. 
Because of the self-similarity, they reduce to 
a set of three ordinary differential equations (ODEs) for $S$, $S'$ and $W$,
together with a constraint equation. (This is discussed in Paper I and also ref.~\cite{cc2000a}).
Another formulation of the field equations, involving  
a set of three ODEs
for $M$, $S$ and $W$ together with a constraint 
equation, is given in ref.~\cite{hm2001}.

We numerically evolve the ODEs
given by Eqs.~(2.27) and (2.28) in Paper I (henceforth referred to as (I.2.27) and (I.2.28))
from small positive $z$ (which corresponds to large physical distance).
We use asymptotically Friedmann solutions in the 
form given by Eqs.~(I.4.1) and (I.4.2), where Eqs.~(I.4.13) and (I.4.14) specify the form of the perturbations.
We set the gauge constants $a_0$ and $b_0$ in Eqs.~(I.3.5) and (I.3.6) to be $1$ for 
the Friedmann solution and
this determines the gauge constants $c_{0}$ and $c_{1}$
through Eqs.~(I.3.11) and (I.3.12).
A single free parameter $A_0$ then characterizes these solutions.
$A_0$ represents the deviation of the energy density from that of the Friedmann universe at spatial infinity, with positive (negative) $A_0$ corresponding to an overdensity (underdensity) there.
We assume $\gamma=1/3$ as a typical example and
numerically integrate the ODEs
using the constraint equation (I.2.31) or (I.2.32) 
to monitor the numerical accuracy.

As shown in Paper I, some of the solutions admit an
extension beyond $z=\infty$ (which corresponds to a finite non-zero physical distance). To assure this, 
we choose suitable dependent variables and then expand these as Taylor series 
with respect to a new independent variable around $z=\pm \infty$.
Paper I specifies the variables which satisfy this requirement for 
asymptotically quasi-Kantowski-Sachs and quasi-static
solutions.
This is our key technique for obtaining the extension of the solution while
retaining good numerical accuracy.
In order to display the solutions, we plot $-1/z$ on the horizontal axis.
The slightly different set of equations
used by Harada and Maeda~\cite{hm2001}
are equivalent to those given in Paper I. 
In fact, we have constructed numerical codes to evolve 
both sets of ODEs
independently and find excellent agreement. 
Our numerical accuracy is checked by the constraint 
equation, which is always satisfied to within a factor of $10^{-7}$.

\begin{figure}[htbp]
\begin{center}
\includegraphics[width=1.0\linewidth]{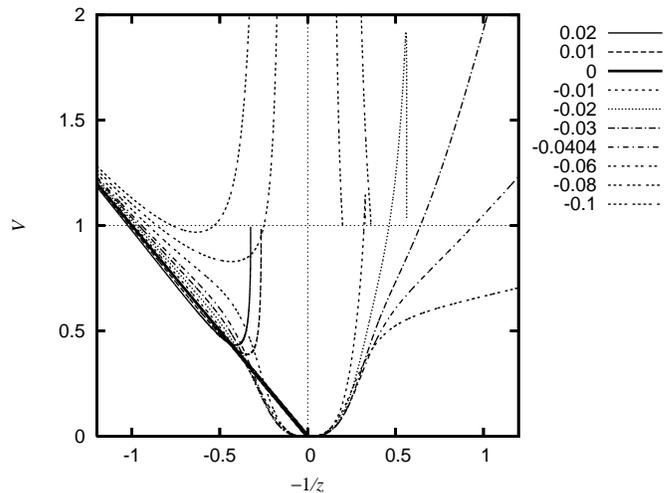}
\caption{\label{V}
$V$ is plotted against $-1/z$ for $\gamma=1/3$.
The thick line corresponds to the exact Friedmann solution ($A_0=0$), which has one similarity horizon.}
\end{center}
\end{figure}

\begin{figure}[htbp]
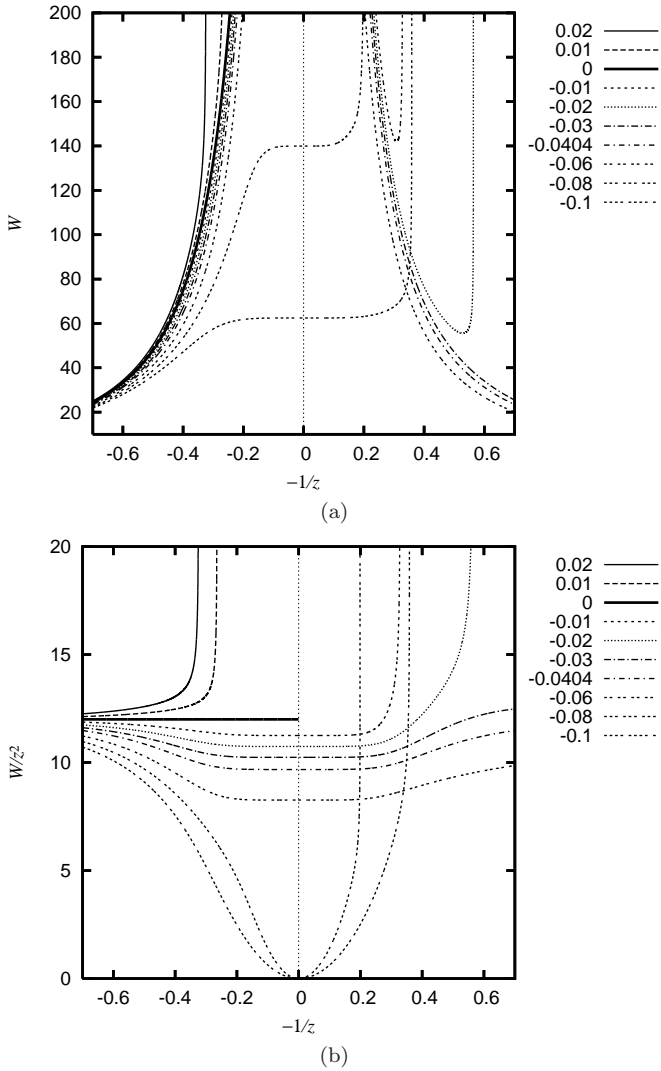

\begin{center}
\subfigure[]{\includegraphics[width=1.0\linewidth]{Fig2a.epsi}}
\subfigure[]{\includegraphics[width=1.0\linewidth]{Fig2b.epsi}}
\caption{\label{W}
(a) $W=8\pi G \mu r^2$ and (b) $W/z^2=8\pi G \mu
 t^2$ are plotted against $-1/z$ for $\gamma=1/3$.
The thick line $W/z^2=12$ corresponds to the exact Friedmann solution ($A_0=0$).
}
\end{center}
\end{figure}

\begin{figure}[htbp]
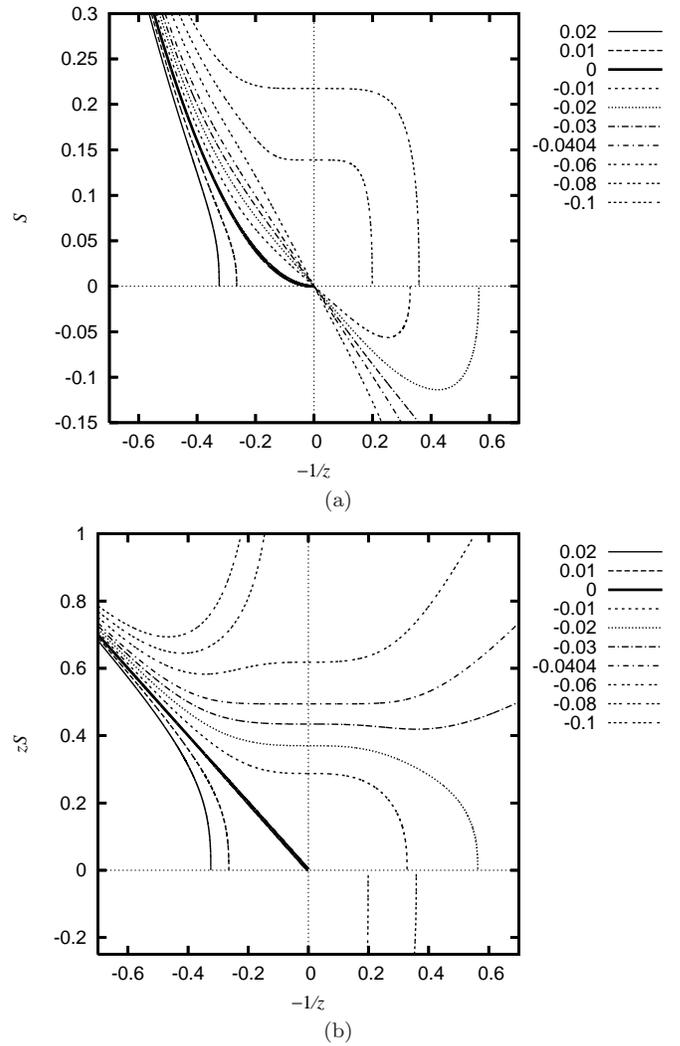

\begin{center}
\subfigure[]{\includegraphics[width=1.0\linewidth]{Fig3a.epsi}}
\subfigure[]{\includegraphics[width=1.0\linewidth]{Fig3b.epsi}}
\caption{\label{S}
(a) $S=R/r$ and (b) $zS=R/t$
are plotted for $\gamma=1/3$. 
The thick line $z^2S=1$ corresponds to the exact Friedmann solution ($A_0=0$).
}
\end{center}
\end{figure}

\begin{figure}[htbp]
\begin{center}
\includegraphics[width=1.0\linewidth]{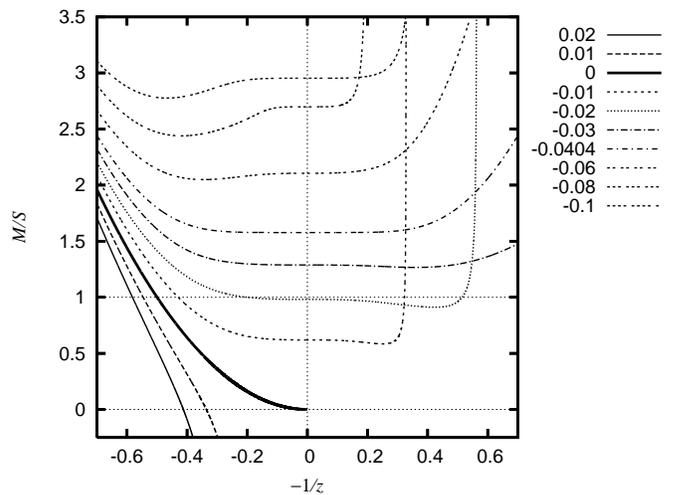}
\caption{\label{MbyS}
$M/S=2Gm/R$ is plotted for $\gamma=1/3$.
The thick line corresponds to the exact Friedmann solution ($A_0=0$).
}
\end{center}

\end{figure}

We show the numerical solutions in Figs.~\ref{V}--\ref{MbyS}.
Although we have used
 many other values for the parameter $A_{0}$,
we only show the results
for $A_{0}=-0.1$, $-0.08$, $-0.06$, $-0.0404$, $-0.03$,
$-0.02$, $-0.01$, 0, 0.01 and 0.02. 
$A_{0}=0$ corresponds to the exact Friedmann solution.

Figure~\ref{V} shows the velocity function $V$.
The thick line corresponds to the exact Friedmann solution and this
has one similarity horizon.
The solutions with positive $A_0$ also have one similarity horizon (where $V=1$) and encounter curvature singularities with negative mass at finite positive $z$, with $V$ converging to $1$ from below.
The solutions with $-0.0253 \lesssim A_0<0$ have two similarity horizons and encounter curvature singularities with positive mass at finite negative $z$, with $V$ converging to $1$ from above.
The solutions with $-0.0780 \lesssim A_0 \lesssim -0.0253$ also have two similarity horizons.
However, of these only the solution with $A_0 \simeq -0.0404$ converges to the asymptotically Friedmann solution as $z \to 0^-$; the other solutions are asymptotically quasi-Friedmann.
The solutions with $A_0 \lesssim -0.0780$ encounter curvature singularities with positive mass at finite negative $z$, with $V$ converging to $1$ from above.
The number of similarity horizons depends on the value of $A_0$:
there are zero, one and two similarity horizons for $A_0 \lesssim -0.108$, $A_0 \simeq -0.108$ and $-0.108 \lesssim A_0 \lesssim -0.0780$, respectively.

Figure~\ref{W} shows the functions $W =8\pi G \mu r^2$ and 
$W/z^2=8\pi G \mu t^2$. 
The thick line $W/z^2=12$ corresponds to the 
exact Friedmann solution ($A_0=0$) and 
has a big-bang singularity as $z \to +\infty$ and $t \to 0$.
$W/z^2$ diverges at finite positive $z$ for solutions with positive $A_0$, corresponding to the negative-mass  curvature singularities seen above.
It also diverges at finite negative $z$ for solutions with $-0.0253 \lesssim A_0<0$, corresponding to the positive-mass singularities.
The solutions with $-0.0780 \lesssim A_0 \lesssim -0.0253$ have no 
singularity in $W/z^2$; instead they converge 
to $12$ as $z \to 0^-$, corresponding to 
the asymptotically quasi-Friedmann solutions.

Figure~\ref{S} shows the functions $S=R/r$ and $zS$.
Note that decreasing (increasing) $S$ as a function of $-1/z$ 
means that the spacetime is expanding (collapsing).
The thick line $z^2S=1$ corresponds to the exact Friedmann solution.
The curvature singularity at finite non-zero $z$ in the solution with 
$A_0\gtrsim -0.0253$ is a central singularity.
There exists a collapsing region near this curvature singularity 
in the solution with $-0.0253 \lesssim A_0<0$. By contrast,
there exists a single local minimum in the profile of $zS$ for the solution with $A_0 \lesssim -0.0253$ and this corresponds to a wormhole throat.
The solution with $-0.0780 \lesssim A_0 \lesssim -0.0253$ has two different spacelike infinities at $z=0^+$ and $0^-$, where $zS$ diverges to $+\infty$.
The solution with $A_0 \lesssim -0.0780$ also has two different spacelike infinities at $z=0^+$ and $z=+\infty$.
In the latter class, $zS$ diverges but $S$ is finite,
 so the solution extends into the negative $z$ (negative $t$) region, where $S$ is negative for positive $R$. 

\begin{figure}[htbp]
\begin{center}
\subfigure[]{\includegraphics[width=1.0\linewidth]{Fig4a.epsi}}
\subfigure[]{\includegraphics[width=1.0\linewidth]{Fig4b.epsi}}
\caption{\label{M}
(a) $M=2Gm/r$ and (b) $zM=2Gm/t$ 
are plotted for $\gamma=1/3$.
The thick line $z^4M=4$ corresponds to the exact Friedmann solution ($A_0=0$).
}
\end{center}
\end{figure}
Figure~\ref{M} shows the functions $M=2Gm/r$
and  $zM=2Gm/t$.
The thick line $z^4M=4$ corresponds to the exact Friedmann solution.
We can see that the curvature singularity at finite non-zero $z$ 
in the solutions with $-0.0253 \lesssim A_0<0$ or $A_0 \lesssim -0.0780$ 
has positive mass, while that in the solution with positive $A_0$ has negative mass.
Note that $zM$ is negative for positive $m$ in the solutions with
$A_0 \lesssim -0.0780$, since negative $z$ corresponds to negative $t$.

Figure~\ref{MbyS} shows the function $M/S=2Gm/R$. 
Note that $M/S>1$ and $M/S<1$ correspond to the trapped and untrapped regions, respectively, and $M/S=1$ in the collapsing (expanding) region corresponds to a future (past) trapping horizon.
The thick line corresponds to the exact Friedmann solution;
this has a single past trapping horizon, corresponding to a cosmological trapping horizon.
The solutions with positive $A_0$ also have a single past trapping horizon.
Those with $A_0 \lesssim -0.0253$ have no trapping horizon and the spacetime is trapped everywhere.
Those with $-0.0253 \lesssim A_0 <0$ have one past trapping horizon and one future trapping horizon.
The latter corresponds to a black hole trapping
horizon.

From these numerical results, we classify 
asymptotically Friedmann solutions as $z \to 0^+$ 
into six classes: (a) the 
flat Friedmann universe ($A_0=0$); (b) a cosmological 
naked singularity ($A_0>0$); (c) 
a cosmological black hole ($\alpha_{1}
\le  A_0<0$); (d) a Friedmann-quasi-Friedmann wormhole ($\alpha_{3}
< A_{0} < \alpha_{1}$ with $A_0 \ne \alpha_{2}$); (e) 
a Friedmann-Friedmann wormhole ($A_0 =\alpha_{2}$); and (f)
a cosmological white hole ($A_0 <\alpha_{3}$). 
For the present choice of 
gauge constants, $\alpha_{1}\simeq -0.0253$, $\alpha_{2}\simeq -0.0404$
and $\alpha_{3}\simeq -0.0780$.

The solutions in class (b) describe a negative mass naked singularity in a Friedmann universe. 
Those in class (c) describe a positive mass black hole in a Friedmann universe.
Those in class (d) contain a wormhole throat connecting a Friedmann universe and a quasi-Friedmann universe.
Class (e) has only one solution and this contains a wormhole throat
connecting two Friedmann universes.
The solutions in class (f) describe a positive mass white hole in a Friedmann universe which contains a wormhole throat connecting a
Friedmann universe and a quasi-static spacelike infinity.
 
The exact flat Friedmann solution, the only solution of class (a), is located at the threshold between classes (b) and (c), while we could not resolve the threshold solution between classes (d) and (f).
The solution of class (e) is obtained by fine-tuning the parameter $A_0$.
We investigate their properties and physically interpret them in the next section.

\section{Physical Interpretation}
\label{sec:physical_interpretation}

\subsection{Exact flat Friedmann universe}
\begin{figure}[htbp]
\begin{center}
\includegraphics[width=0.7\linewidth]{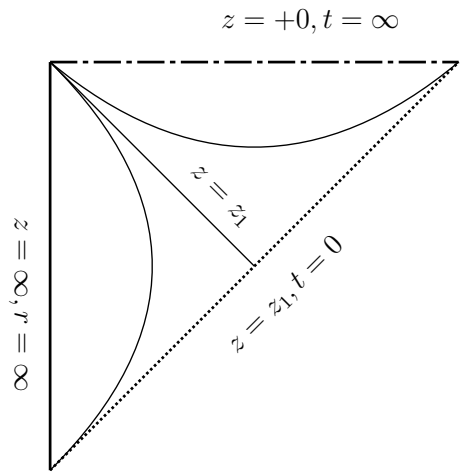}
\caption{\label{fig:FRW} The conformal diagram of the exact flat Friedmann solution. The thin curves and lines denote similarity surfaces, i.e., orbits of $z=$constant.
There is a similarity horizon at $z=z_{1}$ ($>0$).}
\end{center}
\end{figure}
\begin{figure}[htbp] 
\begin{center}
\includegraphics[width=0.8\linewidth]{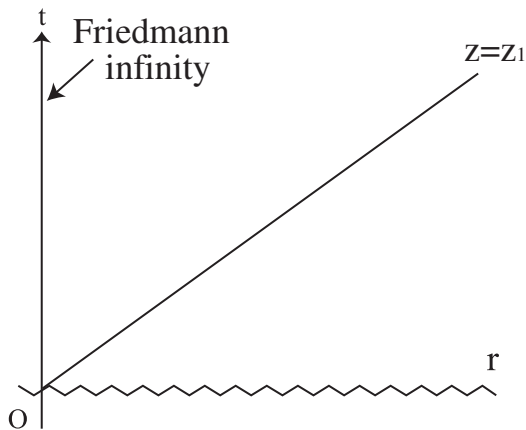}
\caption{\label{tr-FRW}
The $t$-$r$ plane of the exact flat Friedmann solution.
A fluid element moves from the bottom to the top.
$r=0$ and $r \to +\infty$ correspond to spacelike infinity and the physical center ($R=0$), respectively.
$t=0$ corresponds to the big-bang singularity, which is represented by a zig-zag line. 
There is a similarity horizon at $z=z_{1}$ ($>0$), corresponding to a cosmological event horizon.
}
\end{center}

\end{figure}
The solution with $A_0=0$ is the exact flat Friedmann model.
The limit $z \to +\infty$ corresponds to the physical center $R=0$ at constant $t$ or to the 
big-bang singularity at constant $r$. The conformal diagram of this spacetime and the $t$-$r$ plane of the solution are shown in Figs.~\ref{fig:FRW} and \ref{tr-FRW}, respectively. 
There is a similarity horizon at $z=z_{1}$, where $V=1$, corresponding to the cosmological event horizon and the initial singularity is null.

\subsection{Cosmological naked singularity}
\begin{figure}[htbp]
\begin{center}
\includegraphics[width=0.7\linewidth]{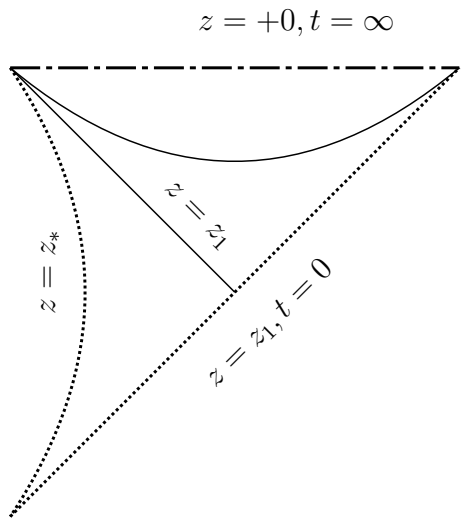}
\caption{\label{fig:CNS} The conformal diagram of the 
cosmological naked singularity solution. There is a similarity 
horizon at $z=z_{1}$ ($>0$). $z=z_{*}$ ($>0$) with
 $0<t<\infty$ corresponds to
a timelike singularity at which the mass is negative.}
\end{center}
\end{figure}
\begin{figure}[htbp] 
\begin{center}
\includegraphics[width=0.8\linewidth]{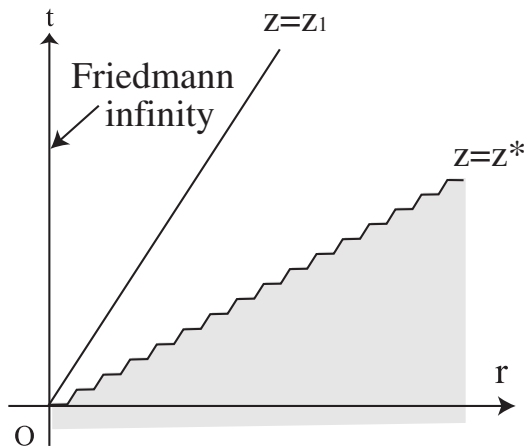}
\caption{\label{tr-CNS}
The $t$-$r$ plane of the cosmological naked singularity solution.
$r=0$ corresponds to spacelike infinity.
There is a similarity horizon at $z=z_{1}$ ($>0$), corresponding to a cosmological event horizon. 
$z=z_{*}$ ($>0$) with $0<t<\infty$ corresponds to a timelike singularity of negative mass and is represented by a zigzag line. 
}
\end{center}
\end{figure}
The solutions with $A_0>0$ represent asymptotically negative-mass 
singularities as $z \to z_{*}$ ($0<z_{*} <\infty$). 
From Fig.~\ref{S}, we see that $R_{,r}>0$ everywhere and also $R_{,t}>0$ (i.e. the
spacetime is expanding), where
a comma denotes a partial derivative.
Figures~\ref{V} and \ref{MbyS} show that there is a cosmological event horizon and a cosmological trapping horizon.
Near the singularity, $V$ approaches 1 from below according to Eq.~(I.4.60) with negative $V_1$.
Figures~\ref{M} and \ref{MbyS} show that the singularity has negative mass
and is in the untrapped region.
This solution describes a naked singularity embedded in 
a Friedmann background. 
The conformal diagram of the spacetime and the $t$-$r$ plane of the solution are shown in Figs.~\ref{fig:CNS} and \ref{tr-CNS}, respectively.
There is a similarity horizon at $z=z_{1}$ and this corresponds to the cosmological event horizon.

\subsection{Cosmological black hole}
The solutions with $\alpha_{1} \le A_0<0$ ($\alpha_{1}\simeq -0.0253$)
are asymptotically quasi-Kantowski-Sachs as $z \to \pm \infty$
and asymptotically positive-mass singular as $z \to z_{*}$ 
($-\infty<z_{*} <0$). 
Because the solutions are asymptotically quasi-Kantowski-Sachs as $z \to +\infty$, they can be extended analytically into the region with negative $z$.
Since the Kantowski-Sachs solution has a curvature singularity at $t=0$, the analytic extension in this case must be interpreted as an extension from the positive $r$ to negative $r$ region.
In the extended region, $S$ and $M$ are negative but the areal radius $R=rS$ and mass $m=rM/(2G)$ are positive because $r$ is negative.

\begin{figure}[htbp]
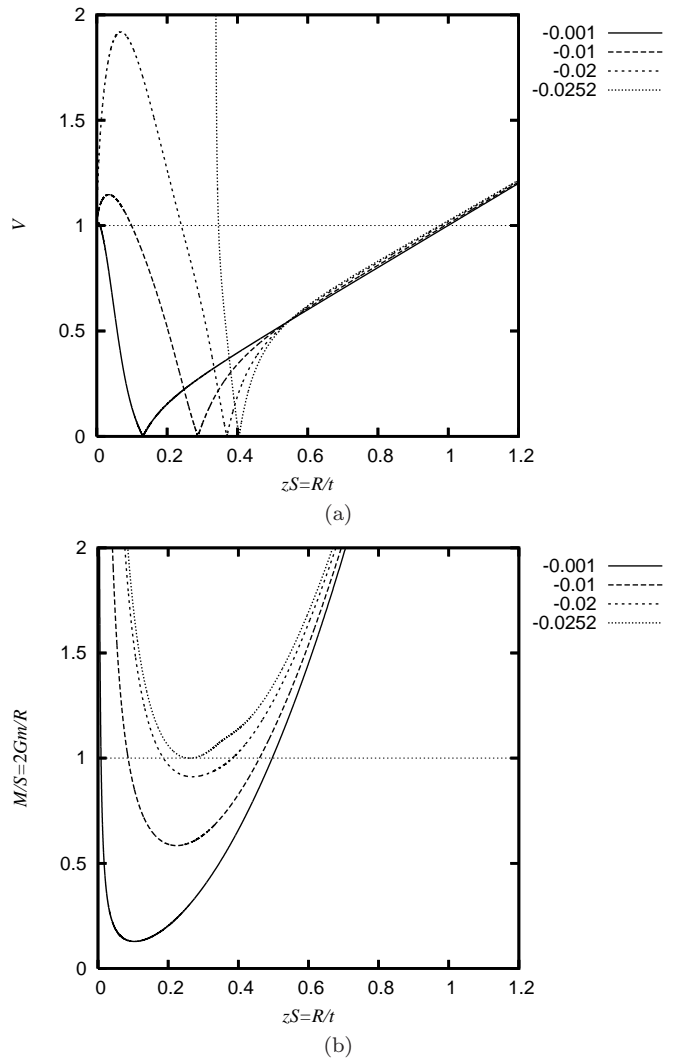

\begin{center}
\subfigure[]{\includegraphics[width=1.0\linewidth]{Fig10a.epsi}}
\subfigure[]{\includegraphics[width=1.0\linewidth]{Fig10b.epsi}}
\caption{\label{fig:CBH-horizon2} 
(a) $V$ and (b) $M/S$ as functions 
of $zS(=R/t)$ for the cosmological 
black hole solutions.
}
\end{center}
\end{figure}

From Fig.~\ref{S}, we find that $R_{,r}>0$ holds everywhere and there exists a collapsing region (with $R_{,t}<0$) near the singularity.
It is seen in Fig.~\ref{MbyS} that there are two trapping horizons, one in the expanding region and the other in the collapsing region.
These are the cosmological and black hole trapping horizons, respectively.
Figure~\ref{fig:CBH-horizon2} shows
that the trapping horizon is degenerate for the critical value $A_0=\alpha_{1}$.
Figures~\ref{V} or \ref{fig:CBH-horizon2} show that there is both a cosmological and black hole event horizons.
For $A_0=-0.01$ and $-0.02$, we see that the black hole 
and cosmological event horizons are in the untrapped and trapped regions, respectively.
Near the singularity, $V$ approaches 1 from above according to Eq.~(I.4.60) with positive $V_1$.
Figures~\ref{M} and \ref{MbyS} show that the singularity has positive mass
and is in the trapped region.

\begin{figure}[htbp]
\begin{center} 
\includegraphics[width=0.8\linewidth]{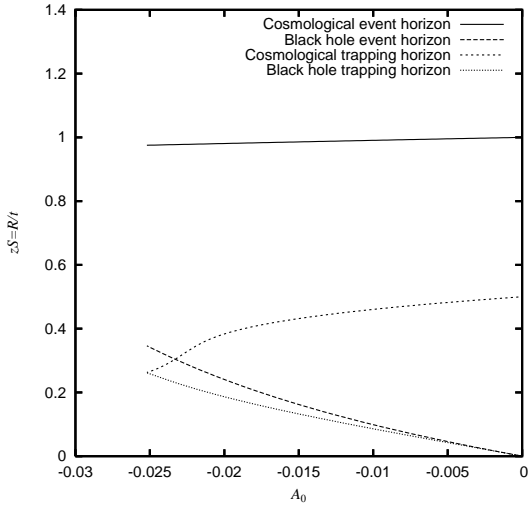}
\caption{\label{horizons}
The value of $zS=R/t$ for various horizons is plotted as a function of the parameter $A_0$ for the cosmological black hole solutions with $\gamma=1/3$. 
}
\end{center}
\end{figure}

The forms of $zS=R/t$ for the various horizons are plotted as functions of the parameter $A_0$  in Fig.~\ref{horizons}.
The ratio of the black hole event horizon size to the cosmological event horizon
size goes from $0$ to $0.36$, while
the ratio of the black hole event horizon size to the 
background Hubble horizon size goes from 0 to 0.70.
On the other hand, the ratio of the cosmological trapping horizon size to 
the black hole trapping horizon size goes from $0$ to roughly $1$.


\begin{figure}[htbp]
\begin{center}
\includegraphics[width=1.0\linewidth]{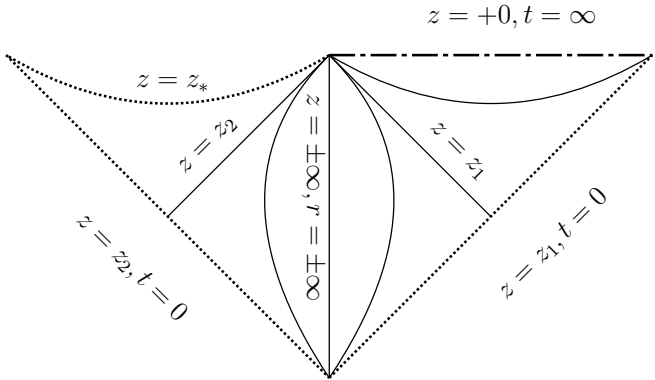}
\caption{\label{fig:CBH} The conformal diagram of the 
cosmological black hole solution. There are two 
similarity horizons at $z=z_{1}$ and $z=z_{2}$ ($z_{2}<0<z_{1}$).
$z=z_{1}$ with $0<t<\infty$ is the cosmological event horizon, while $z=z_{2}$ with $0<t<\infty$ is the black hole event horizon.
$z=z_{*}$ ($<0$) with $0<t<\infty$ gives a spacelike singularity with positive mass.
}
\end{center}
\end{figure}

\begin{figure}[htbp] 
\begin{center}
\includegraphics[width=1.0\linewidth]{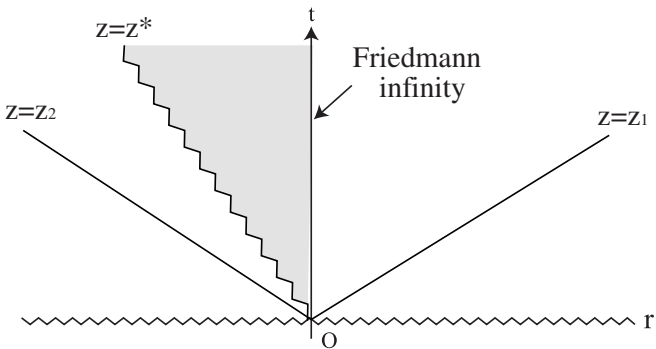}
\caption{\label{tr-CBH}
The $t$-$r$ plane of the cosmological black hole solution.
$r=0^+$ corresponds to spacelike infinity.
}
\end{center}

\end{figure}

This solution describes a black hole in a Friedmann background. 
Figures~\ref{fig:CBH} and \ref{tr-CBH} show the conformal diagram and $t$-$r$ plane for the cosmological black hole solution. 
We can see that the initial singularity is null and that there exists both a null and spacelike portion of the black hole singularity.
There are two similarity horizons 
at $z_{1}>0$ and $z_{2}<0$. These
correspond to the cosmological and black hole event horizons, respectively.

\subsection{Friedmann-quasi-Friedmann wormhole}
\begin{figure}[htbp]
\begin{center}
\subfigure[]{\includegraphics[width=1.0\linewidth]{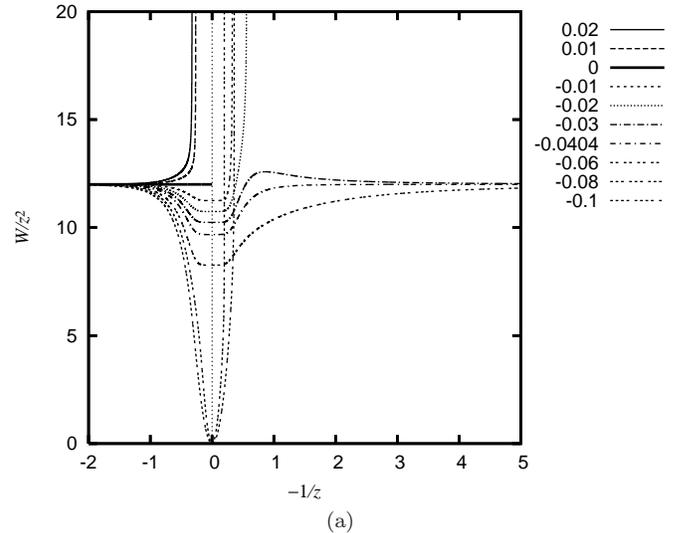}}
\subfigure[]{\includegraphics[width=1.0\linewidth]{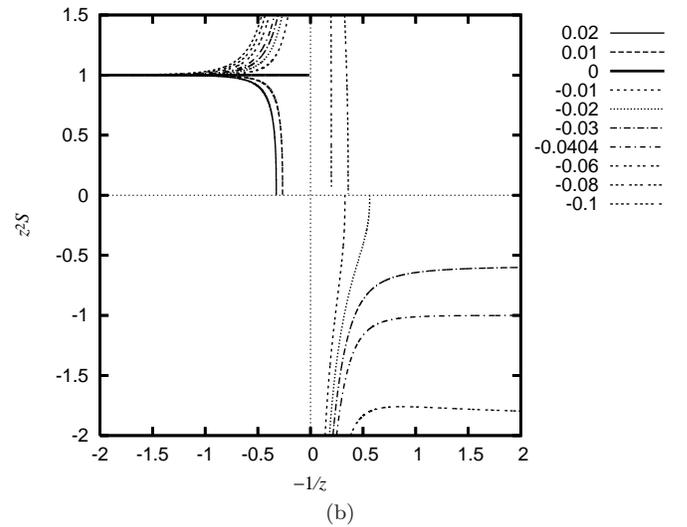}}
\caption{\label{Fconv}
(a) $W/z^2$ and (b) $z^2S$ are plotted for $\gamma=1/3$.
}
\end{center}
\end{figure}

The solutions with $\alpha_{3} < A_0 < \alpha_{1}$
($\alpha_{3}\simeq -0.0780$) are also asymptotically quasi-Kantowski-Sachs as $z \to \pm\infty$ and the form of these solutions in the extended region can again be obtained numerically.
Figure~\ref{Fconv} shows that $W/z^2 \to 12$ 
and $z^2S \to$ constant as $z \to 0^-$. In the
asymptotically Friedmann case ($A_0 = \alpha_{2} \simeq -0.0404$) the constant is $-1$. Otherwise the solutions are asymptotically
quasi-Friedmann as $z \to 0^-$ and the constant is different from $-1$.

From the profile of $zS$ in Fig.~\ref{S}, we see that a solution in this class has a single throat at $z=z_{\rm t}$, characterized by $R_{,r}=0$ and $R_{,r}<(>) 0$ for $-1/z <(>) -1/z_{\rm t}$, i.e. the areal radius $R$ has a minimum on the spacelike hypersurface with constant $t$.
As a result, the spacetime represents a dynamical wormhole with a throat connecting a Friedmann universe to another Friedmann or quasi-Friedmann universe.
With $A_0 =\alpha_{2}$, the solution has reflection symmetry, so the wormhole throat connects two exact Friedmann universes and is located at $z=\pm \infty$.
Fine-tuning of $A_0$ is required because the asymptotically
Friedmann solution given by Eqs.~(I.4.13) and (I.4.14) is non-generic among the solutions of the linearized equations for $A$, $A'$, $B'$ and $B''$ in the region where $V^2 \to \infty$
(see Paper I).

This wormhole universe is born from the initial singularity at $t=0$, around which the spacetime for fixed $r$ is described by the quasi-Kantowski-Sachs solution.
The profile of $S$ in Fig.~\ref{S} shows that the spacetime is expanding everywhere
and Fig.~\ref{M} shows that it has positive mass.
From Figs.~\ref{V} and \ref{MbyS}, there are two similarity horizons, corresponding to two cosmological event horizons, but there is no trapping horizon.

\begin{figure}[htbp]
\begin{center}
\includegraphics[width=1.0\linewidth]{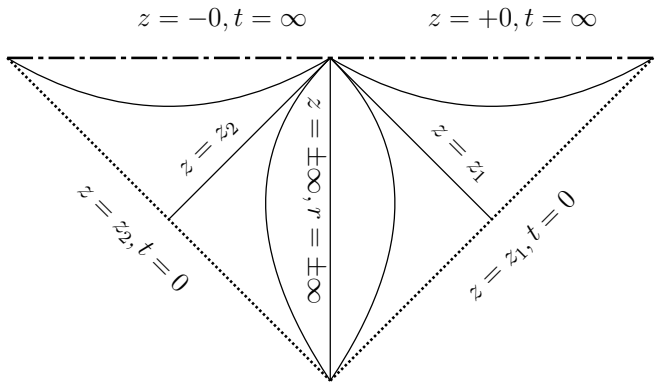}
\caption{\label{fig:FFWH} The conformal diagram of the 
Friedmann-quasi-Friedmann wormhole solution. There are two 
similarity horizons, $z=z_{1}$ and $z=z_{2}$ ($z_{2}<0<z_{1}$), 
both corresponding to cosmological event horizons.
$z=0^+$ and $z=0^-$ give two distinct null infinities, described by Friedmann and quasi-Friedmann asymptotics, respectively. 
$z=\pm\infty$, however, 
is described by the Kantowski-Sachs asymptotic
as $t$ increases from 0 to $\infty$.}
\end{center}
\end{figure}

\begin{figure}[htbp]
\begin{center} 
\includegraphics[width=1.0\linewidth]{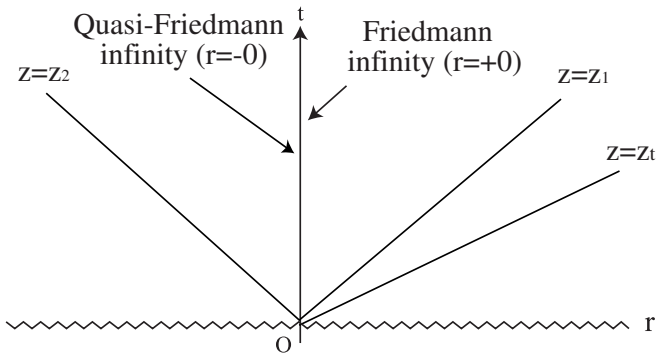}
\caption{\label{tr-FQFWormhole}
The $t$-$r$ plane of the Friedmann-quasi-Friedmann wormhole solution.
We show the solution with a wormhole throat at $z=z_{\rm t}$ in the positive $z$ region.
$r=0^+$ and $0^-$ correspond to the Friedmann and quasi-Friedmann
 infinities, respectively.
$t=0$ corresponds to the big-bang singularity, represented by the quasi-Kantowski-Sachs solution. 
}
\end{center}
\end{figure}

Figures~\ref{fig:FFWH} and \ref{tr-FQFWormhole} show the conformal diagram and $t$-$r$ plane for the Friedmann-quasi-Friedmann wormhole solution. 
We can see that the initial singularity is null and that there are two parts
of future null infinity, both being spacelike and 
having Friedmann and quasi-Friedmann asymptotics.
There are two similarity horizons at $z_{1}>0$ and $z_{2}<0$, both corresponding to cosmological event horizons. Figure~\ref{MbyS} shows that the spacetime is trapped everywhere, 
so there is no trapping horizon.
This explicit example shows 
that the wormhole definitions
of Hochberg and Visser~\cite{hv1997} or Hayward~\cite{hayward1999}
are too restrictive.


\subsection{Cosmological white hole}
The solutions with $A_0 < \alpha_{3}$ are asymptotically quasi-static
as $z \to \pm \infty$, 
so that $S=R/r$ converges to a constant there
(see Fig.~\ref{S}). We can therefore extend them analytically into the negative $z$ region, which corresponds to negative $t$.

The variables $S$ and $W$ can be used to obtain the solutions in the extended region because they are finite in the asymptotically quasi-static solutions at $z=\pm\infty$.
In the extended region, $zS$ and $zM$ are negative but the areal radius $R=rS$ and mass $m=rM/(2G)$ are positive because $t$ is negative.

The profile of $zS=R/t$ in Fig.~\ref{S} shows that its minimum value corresponds to a throat connecting a flat Friedmann infinity and a quasi-static infinity, so this describes a dynamical wormhole.
The spacetime is expanding everywhere and
has positive mass from Fig.~\ref{M}.
Figure~\ref{MbyS} shows that it is trapped everywhere, so there is no trapping horizon.

As seen in Figs.~\ref{W} and \ref{S}, there is a central curvature singularity at $z=z_{*}$ in the negative $z$ region and the mass is positive there from Fig.~\ref{M}.
This 
corresponds to the singular positive-mass asymptote discussed in Paper I.
Near the singularity, $V$ 
is approximated by Eq.~(I.4.60) with positive $V_1$ and approaches $1$ from Fig.~\ref{V}.
The singularity is spacelike because of the positive mass and resembles a Schwarzschild white hole singularity (although there is no trapping horizon).
This singularity is non-simultaneous, so it differs from the big-bang singularity.
On the other hand, the singularity at $t=0$ with $z=z_1$ is massless and does correspond to the big-bang singularity.
The solution describes a white hole in the Friedmann universe.

As seen in Fig.~\ref{V}, the solutions can be classified into
three types, according to whether they have no similarity horizon
($A_0 \le  \alpha_{4}$), one degenerate similarity horizon ($A_0 =
\alpha_{4}$) or two non-degenerate similarity horizons ($\alpha_{4} < A_0
< \alpha_{3}$), where $\alpha_{4}\simeq -0.108$.
The degenerate case is obtained by fine-tuning $A_0$.

\begin{figure}[htbp]
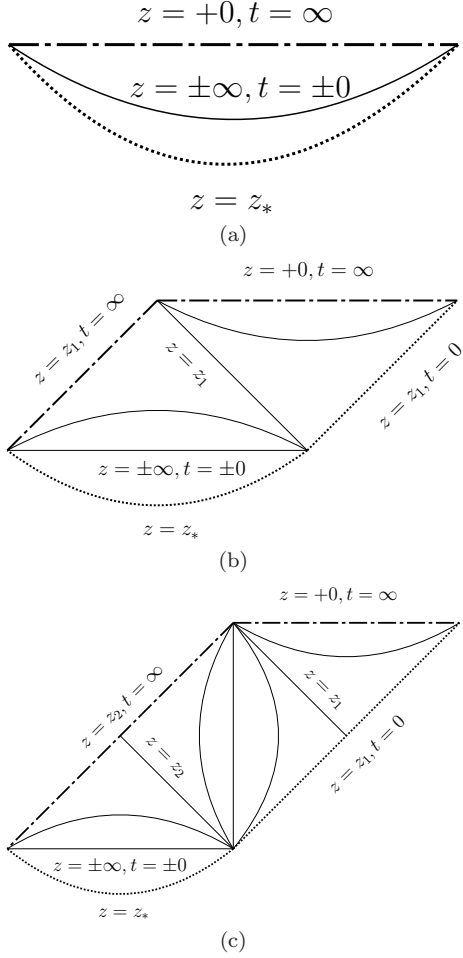

\begin{center}
\subfigure[]{\includegraphics[width=0.7\linewidth]{Fig17a.epsi}}
\subfigure[]{\includegraphics[width=0.7\linewidth]{Fig17b.epsi}}
\subfigure[]{\includegraphics[width=0.7\linewidth]{Fig17c.epsi}}
\caption{\label{fig:FQSWH} The conformal diagrams of the 
cosmological white hole solutions with: 
(a) no similarity horizon;
(b) one degenerate similarity horizon $z=z_{1}$ ($>0$);
and (c) two distinct similarity horizons $z=z_{1}$ and $z=z_{2}$ ($0<z_{1}<z_{2}<\infty$). For each case, 
$z=z_{*}$ ($<0$) is 
a spacelike singularity with positive mass. 
In (b), $z=z_{1}$ with $t=\infty$ corresponds to future null infinity.
In (c), $z=z_{2}$ with $t=\infty$ corresponds to future null infinity.
}
\end{center}
\end{figure}

\begin{figure}[htbp] 
\begin{center}
\includegraphics[width=0.7\linewidth]{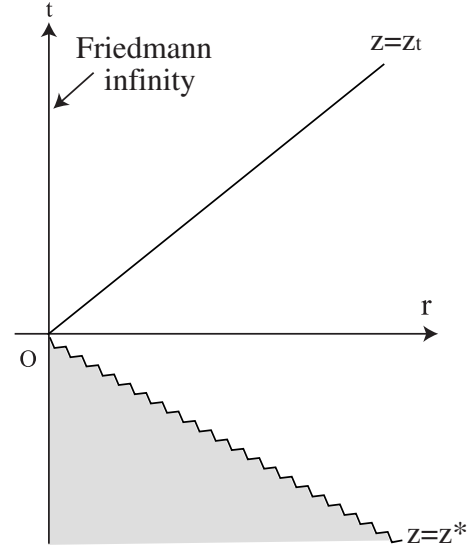}
\caption{\label{tr-FQSWormhole}
The $t$-$r$ plane of the cosmological white hole solution in the case without a similarity horizon.
$r=0^+$ and $+\infty$ correspond to the Friedmann and quasi-static infinities, respectively.
There is a wormhole throat at $z=z_{\rm t}$ in the positive $z$ region.
$z=z_{*}$ ($<0$) with $0<t<\infty$ corresponds to a spacelike singularity with positive mass, represented by a zigzag line. 
}
\end{center}
\end{figure}

Figures~\ref{fig:FQSWH} and \ref{tr-FQSWormhole} show the conformal diagram and $t$-$r$ plane for these three types of Friedmann-quasi-static wormhole solutions. 
In the case with similarity horizon(s), 
the singularity has both spacelike and null 
portions, which correspond to the white hole and initial singularities, respectively, and there are two parts of future null infinity. 
One is spacelike and described by the Friedmann asymptote and the other is null.
In the case with no similarity horizon, the singularity
and future null infinity are both spacelike.


\begin{figure}[htbp]
\begin{center}
\includegraphics[width=0.7\linewidth]{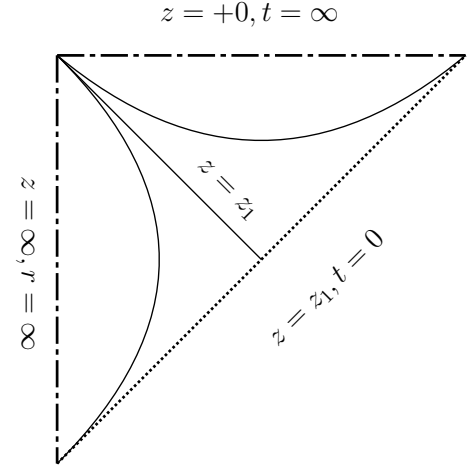}
\caption{\label{fig:FACV} The fictitious conformal diagram of the 
Friedmann-constant-velocity wormhole solution. 
There is a similarity horizon at $z=z_{1}$ ($>0$), corresponding to the cosmological event horizon.
$z=\infty$ with $0<t<\infty$ corresponds to a timelike boundary of spacetime at infinity.
}
\end{center}
\end{figure}

\bigskip 

In closing this section, a comment should be made about what is meant by the term ``wormhole''. It is sometimes assumed that wormholes require negative energy density and there are certainly theorems to this effect~\cite{visser,hv1997,hayward1999}. 
Our solutions do not require negative energy density but this does not conflict with these theorems because we have a different definition of a wormhole throat.

The theorems claiming the violation of the null energy condition at a wormhole throat~\cite{hv1997,hayward1999} define the throat on a null hypersurface.
In this case, the throat coincides with a trapping horizon.
In the present paper we define a wormhole throat on a spacelike hypersurface. 
In fact, our wormhole solutions do not have a trapping horizon at all, so 
they are not wormholes in the sense of Hochberg and Visser~\cite{hv1997} or Hayward~\cite{hayward1999} even though they possess two distinct infinities. 
We discuss this problem in a more general framework in a separate paper~\cite{wormhole}.

Here we also consider the threshold between the Friedmann-quasi-Friedmann
solutions and the cosmological white hole solutions. Strictly speaking, the
threshold solution with the exact parameter value $A_{0}=\alpha_{3}$
cannot be achieved numerically. Among the asymptotic solutions obtained in
Paper I, the constant-velocity asymptote will be the most plausible candidate for this
threshold solution. The resulting conformal diagram would be given by Fig.~\ref{fig:FACV},
admitting a null ray traveling from the constant velocity infinity ($z\to +\infty,
r\to +\infty$) to the Friedmann infinity ($z\to +0, t\to \infty$). However, this
possibility is clearly excluded because this diagram contains a traversable
dynamical wormhole, which is prohibited by the discussion in Hochberg and Visser~\cite{hv1997} or Hayward~\cite{hayward1999}.

\section{Summary and discussion} 
\label{sec:summary_discussion}
\begin{table}[h]
\begin{center}
\caption{\label{table:summary_solutions} Self-similar solutions with $\gamma =1/3$ which are
 asymptotic to a Friedmann universe at large distance.
The numerical values are given by $\alpha_1 \simeq -0.0253$, $\alpha_2
 \simeq -0.0404$, $\alpha_3 \simeq -0.0780$ and $\alpha_{4}\simeq
 -0.108$ for the choice of the gauge constants $a_0=b_0=1$.
The terms similarity horizon, trapping horizon, Friedmann, quasi-static,
quasi-Kantowski-Sachs,  positive-mass singular,
negative-mass singular are abbreviated as SH, TH, F, QS,
QKS, PMS and NMS, respectively. The last two columns give the number of similarity and trapping horizons, with 1(d) indicating one degenerate horizon. 
We note that the spacetime for $A_0=\alpha_3$ has not been determined.
}
\begin{tabular}{|c||c|c|c|c|c|c|}
\hline
$A_0$ & Spacetime & Asymptote & \# SH
  & \# TH \\ \hline \hline
 & Naked singularity & F-NMS & 1 & 1 \\ \hline
0 & Friedmann universe & F & 1 & 1 \\ \hline
 & Black hole & F-QKS-PMS & 2 & 2\\ \hline
$\alpha_1$ & Black hole & F-QKS-PMS & 2 & 1 (d)\\ \hline
 & Wormhole  & F-QKS-QF & 2 & 0\\ \hline
$\alpha_2$ & Wormhole  & F-QKS-F & 2 & 0 \\ \hline
 & Wormhole & F-QKS-QF & 2 & 0 \\ \hline
$\alpha_3$ & ? & ? & ? & ? \\ \hline
 & White hole & F-QS-PMS & 2 & 0\\ \hline 
$\alpha_4$ & White hole & F-QS-PMS & 1(d) & 0\\ \hline
 & White hole & F-QS-PMS & 0 & 0
 \\ \hline 
\end{tabular}
\end{center}
\end{table} 
We have numerically investigated spherically symmetric self-similar solutions for a perfect fluid with $p=(\gamma-1)\mu$ ($0<\gamma<2/3$) which are properly asymptotic to the flat Friedmann model at large distances,
in the sense that there is no solid angle deficit.
We have integrated the field equations for the self-similar solutions and 
interpreted the solutions using the asymptotic analysis of Paper I.
The results are summarized in Table I.

We have seen that there is a class of asymptotically Friedmann self-similar black hole solutions 
and this contrasts to the situation with positive pressure fluids or scalar fields, where there are no such solutions.
This suggests that self-similar growth of PBHs is possible in an
inflationary flat Friedmann universe with dark energy, which may support the claim in ref.~\cite{bm2002}.
This can be explained by the fact that 
the pressure gradient of dark energy is in the opposite
direction to the density gradient. It therefore gives an attractive force 
which pulls the matter inwards, whereas the gravitational force is 
repulsive.
So if the black hole is small and the density gradient steep,
it can accrete the surrounding mass effectively. 

For $\gamma=1/3$, we have found numerically that the ratio of the size
of the black hole event horizon to the Hubble length is between $0$ and $0.70$.
This means that the size of such a self-similar 
cosmological black hole has an upper limit  but it can be 
arbitrarily small, so
 black holes which grow as fast as the universe cannot be too large.
It should be stressed that there is a one-parameter family 
of self-similar cosmological black hole solutions, so 
we do not have to fine-tune the 
mass of the black hole to get 
self-similar growth.
This means that self-similar black hole growth is very plausible 
in a universe with dark energy of this kind.
This is in contrast to the usual Newtonian argument
for a positive pressure fluid~\cite{zn1967,hc2005c}.

In addition to the black hole solutions, we have 
found self-similar solutions which represent a Friedmann universe connected to either another Friedmann universe or some other cosmological model.
They are interpreted as self-similar cosmological white hole or wormhole solutions.
These are classified into three types, according to whether the other infinity is 
Friedmann, quasi-Friedmann, or quasi-static, among which the last type is interpreted as a white hole in a Friedmann background.
These solutions provide intriguing examples of dynamical wormholes.
Since they are cosmological solutions, they are not asymptotically flat and they do not have a regular past null infinity. Rather they are asymptotic to models which are expanding at spacelike infinity and have an initial singularity. 

In such situations a wormhole throat should be locally defined as a two-sphere of minimal area on the spacelike hypersurfaces rather than a trapping horizon.
According to the conventional definition of traversability, none of our numerical solutions represents a traversable wormhole because of the absence of a past null infinity. 
The strong energy condition is violated there but the null, dominant and weak energy conditions are still all satisfied.
Our wormhole solutions do not have a trapping horizon: because there is no violation
of the null energy condition, they are not dynamical wormholes in the sense of Hochberg and Visser~\cite{hv1997} or Hayward~\cite{hayward1999}.

Finally, we note that the homothetic (rather than comoving) approach is another powerful
method of studying self-similar solutions.
In the positive pressure case, the homothetic and comoving approaches are 
complementary, which leads to a comprehensive understanding 
of self-similar solutions~\cite{cc2000a,gnu1998,ccgnu2001}.
A dynamical systems approach in the negative pressure case should also
 shed light on various aspects of the problem.


\acknowledgments
The authors would like to thank S.A.~Hayward, P.~Ivanov, H.~Kodama, H.~Koyama,
M.~Siino and T.~Tanaka for useful comments. 
HM and TH are supported by the Grant-in-Aid for Scientific
Research Fund of the Ministry of Education, Culture, Sports, Science
and Technology, Japan (Young Scientists (B) 18740162 (HM) and 18740144 (TH)). BJC thanks the Research Center for the Early Universe at the University of Tokyo for hospitality received during this work.
HM was also supported by the Grant No. 1071125 from FONDECYT (Chile).
CECS is funded in part by an institutional grant from Millennium Science
Initiative, Chile, and the generous support to CECS from Empresas CMPC 
is
gratefully acknowledged.


\end{document}